\documentstyle[11pt]{article}
\begin{document}
\topmargin -15mm
\textheight 225mm
\baselineskip14pt
 
\pagestyle{empty}
\rightline{UG-1/96}
\rightline{March 1996}
\rightline{hep-th/yymmddd}
\vspace{2truecm}
\centerline{\bf $D$--Branes and $T$--Duality}
\vspace{2truecm}
\centerline{E. BERGSHOEFF AND M. DE ROO}
\vspace{.2truecm}
\centerline{\it Institute for Theoretical Physics}
\centerline{\it University of Groningen, Nijenborgh 4}
\centerline{\it 9747 AG Groningen}
\centerline{\it The Netherlands}
\vspace{3truecm}
\centerline{ABSTRACT}
\vspace{.5truecm}
We show how the $T$--duality between $D$--branes is realized 
(i) on $p$--brane solutions $(p=0,\cdots ,9)$ of IIA/IIB supergravity 
and (ii) on the $D$--brane actions ($p=0,\cdots ,3)$ that act as 
source terms for the $p$--brane solutions. We point out that 
the presence of a cosmological
constant in the IIA theory leads, by the requirement of gauge
invariance, to a topological mass term for the worldvolume
gauge field in the 2--brane case.

\vfill\eject
\pagestyle{plain}
 
\noindent{\bf 1. Introduction}
 
\vspace{.5cm}

Recent developments in string theory have shown that $p$--brane
solutions and duality symmetries play an important role in understanding
the nonperturbative behaviour of the theory. An important example
of a duality symmetry is the $T$--duality \cite{Gi1}
which states
that a string compactified on a torus with radius $R$ is equivalent
to a string compactified on a torus with radius $\alpha^\prime/R$
where $\alpha^\prime$ is the inverse string tension.

It turns out that the $p$--brane solutions whose charge are carried by a 
RR (Ramond/Ramond) gauge field of the type II supergravity theories
have a natural place within open string theory as $D$--branes \cite{Pol1}.
The relation is established via the requirement that the endpoints of the 
open string are constrained to live on the $p+1$--dimensional worldvolume of 
the Dirichlet $p$--brane. Such a (ten--dimensional) open string state 
is described by Dirichlet boundary
conditions for the $9-p$ transverse directions and Neumann boundary
conditions for the $p+1$ worldvolume directions. Since under $T$--duality
Dirichlet and Neumann boundary conditions are interchanged it follows that
all Dirichlet $p$--branes $(p=0,\cdots ,9)$ are $T$-dual versions
of each other. A discussion of how this $T$--duality between
$D$--branes arises in string theory can be found in the recent 
review article \cite{Pol2}. 

Since all $D$--branes are $T$--dual to each other it is natural to expect
that this $T$--duality is also realized on the underlying $p$--brane
solutions of the IIA/IIB supergravity theories. Furthermore, the $T$--duality
should also be realized on the Dirichlet $p$--brane actions which act
 as source terms of the $p$--brane solutions. It is the purpose
of this letter to give the details of this $T$--duality between
 Dirichlet $p$--brane solutions and their source terms and to point
out a few subtleties that occur in establishing $T$--duality.

\vspace{.5cm}

\noindent{\bf 2.\ $D$--Brane Solutions}

\vspace{.5cm}

We first consider the $T$--duality between the Dirichlet $p$--brane 
solutions. In \cite{Be1} it was pointed out that the $T$--duality between the
 RR $p$--brane solutions of $D=10$ IIA/IIB supergravity, 
with $0 \le p \le 9$, is an almost immediate consequence of
 the $T$--duality rules of \cite{Be2} which are a generalization to
curved background of the type II $T$--duality rules of \cite{Da1,Di1}. 
The relationship between the 7--brane and 8--brane
solutions turns out to be more subtle and it involves the introduction
of modified, so-called {\sl massive} $T$--duality rules \cite{Be1}.

We will be concerned with a class of solutions whose metric and
 dilaton ($\sigma$) for all values of $p$ ($0\le p \le 9)$ is given
by 

\begin{eqnarray}
\label{gpbrane}
 ds^2 &=& H^{-{1\over 2}} d^2s_{p+1} + H^{{1\over 2}}ds^2_{9-p}
\, ,\nonumber\\
e^{-2\sigma} &=& H^{{1\over 2}(p-3)}\, ,
\end{eqnarray}
where $d^2 s_{p+1}$ is the Minkowski $(p+1)$--metric on the
 worldvolume
and $ds^2 s_{9-p}$ is the Minkowski $(9-p)$--metric on the 
transverse space. The function $H$ only depends on the transverse
 coordinates and is harmonic with respect to these variables. The dilaton
$\sigma$ is equal to the dilaton $\phi$ ($\varphi$) of IIA (IIB)
supergravity depending on whether $p$ is even or odd:

\begin{eqnarray}
\label{sigma}
\sigma &=& \phi \ ({\rm IIA\ dilaton})\hskip 1.5truecm p\ {\rm
 even}\, ,\nonumber\\
\sigma &=& \varphi \ ({\rm IIB\ dilaton}) \hskip 1.5truecm p\ {\rm odd}\, .
\end{eqnarray}
We use here the same notation and
conventions as \cite{Be1,Be2}. The fact that
the RR p-brane solutions for $0\le p \le 6$ are of the form
(\ref{gpbrane}) was shown in \cite{To1}. Each of the above RR p-brane
solutions has a nontrivial RR (electric or magnetic) charge carried by
one of the RR gauge fields. We will first discuss the duality of
the metric and dilaton and after that the
behaviour under duality of the corresponding RR gauge fields.

It is not too difficult to see that the class of metrics and dilatons
given in (\ref{gpbrane}) are
 transformed into each other under $T$--duality. Although the
$T$--duality rules are complicated for general backgrounds
they simplify for the class of solutions (\ref{gpbrane})
which have a diagonal metric and a vanishing NS/NS 2--form gauge field. The
relevant rules of the metric and dilaton are in this case given 
by\footnote{We only give here the rules that lead from a given
IIA solution to a dual IIB solution. The inverse rules
can be used to construct a dual
IIA solution out of a given IIB solution.}

\begin{eqnarray}
\label{duality}
j_{\mu\nu} &=& g_{\mu\nu}\, ,\nonumber\\
j_{xx} &=& 1/ g_{xx}\, ,\\
e^{-2\varphi} &=& e^{-2\phi} |g_{xx}|\, .\nonumber
\end{eqnarray}
Here $g\ (j)$ is the IIA\ (IIB) metric. The coordinate $x$ refers to the 
 direction over which the duality is
 performed and the $\mu$ index labels the nine remaining directions.

Starting with the 0--brane, or particle solution,
 one first dualizes over one
of the transverse directions, say $x^1$, thereby assuming that the
function $H$ of the 0--brane solution is independent of $x^1$ and
is harmonic over the remaining eight transverse directions
$x^2, \cdots x^9$. Applying the
duality rules (\ref{duality}) with respect to the $x^1$ direction
 leads to the 1--brane solution given in (\ref{gpbrane}).
Next, one assumes that the harmonic function corresponding to the
 1-brane solution is independent of one of its transverse
 directions, say $x^2$. Dualizing over $x^2$ then leads to the 
2-brane solution. This process is repeated until one reaches
the 8-brane solution. At this point the transverse
space has become one--dimensional so that the harmonic function $H$
only depends on one variable, say $x^9$. In the last step one assumes that
the dependence on $x^9$ disappears so that we end up with a 9-brane
solution that is given by flat 10--dimensional Minkowski spacetime.

Next, we consider the behaviour of the RR gauge fields under $T$--duality.
The RR gauge fields of the type IIA theory are given by
$\{A_\mu^{(1)}, C_{\mu\nu\rho}\}$ while those of the type IIB
theory are given by $\{D_{\mu\nu\rho\sigma}, {\cal B}_{\mu\nu}^{(2)},
\ell\}$.
Again, the duality rules for a general background are complicated
but simplify for the
class of solutions we are considering. They are given by:

\begin{eqnarray}
\label{duality2}
D_{\mu\nu\rho x} &=& {3\over 8} C_{\mu\nu\rho}\, ,\nonumber\\
{\cal B}_{\mu\nu}^{(2)} &=& {3\over 2}C_{\mu\nu x}\, ,\\
{\cal B}^{(2)}_{x \mu} &=& - A_\mu^{(1)}\, ,\nonumber\\
\ell &=& A_x^{(1)} + m x\, .\nonumber
\end{eqnarray}
Here $m$ is the mass parameter of massive IIA
supergravity \cite{Ro1}.

Two remarks are in order. First of all, we have only given the
duality rule of $D_{\mu\nu\rho x}$. The duality rule for
$D_{\mu\nu\rho\sigma}$ follows from the self-duality condition 

\begin{equation}
\label{selfduality}
F(D)_{\mu_1\cdots \mu_5} = {1\over 120}
{1\over \sqrt {-j}}j_{\mu_1\nu_1}\dots
j_{\mu_5\nu_5}\epsilon^{\nu_1\cdots \nu_5\rho_1\cdots \rho_5}
F(D)_{\rho_1\cdots \rho_5}\, ,
\end{equation}
where $F(D) = d D$ is the curvature of $D$ for vanishing NS/NS 2--form gauge
 fields.
From this it follows that we cannot determine the duality rule for
$D_{\mu\nu\rho\sigma}$ but only of its curvature $F(D)_{\mu_1
\cdots \mu_5}$. 
Secondly, the mass parameter in the last rule of (\ref{duality2}) 
will only play a role in
establishing the duality between the 7--brane and 8--brane solutions,
as has been explained in \cite{Be1} and will be discussed further below.

Using the duality rules (\ref{duality2}) for the RR gauge fields, it
is straightforward to get the expressions of the RR gauge fields (or
curvatures) of
{\it all} $p$--brane solutions, starting from the particle solution.
We find the following expressions:

\begin{equation}
\label{RRgf}
\begin{array}{rclrcl}
p&=&0: & A_0^{(1)} &=& H^{-1}\, , \\
p&=&1:& {\cal B}_{01}^{(2)} &=& H^{-1}\, \\
p&=&2:& C_{012} &=& {2\over 3} H^{-1}\, \\
p&=&3:& D_{0123} &=& {1\over 4} H^{-1}\, \ \ {\rm or}
\\
&&&F(D)_{ijklm} &=& {1\over 20}
\epsilon^{ijklmn} H^2\partial_n H^{-1}\, ,\\
p&=&4:& 
G_{ijkl} &=& {1\over 6}\epsilon^{ijklm}H^2\partial_m H^{-1}\, \\
p&=&5: &{\cal H}^{(2)}_{ijk} &=& -{1\over 3}
\epsilon^{ijkl}H^2
\partial_l H^{-1}\, ,\\
p&=&6:& F^{(1)}_{ij} &=& -\epsilon^{ijk}H^2\partial_k 
H^{-1}\, , \\
p&=&7:& \partial_i\ell &=& \epsilon^{ij}H^2\partial_j
 H^{-1}\, ,\\
p&=&8:& m &=& H^\prime\, ,\\
p&=&9:& H &=& 1\, .\\
\end{array}
\end{equation}
Here the $0,1,2,\cdots$ indices refer to the $(p+1)$--dimensional worldvolume
 directions and $i$ denotes the $(9-p)$ transverse directions. 
Furthermore, $F(D)=d D, G=dC, {\cal H}^{(2)}=d{\cal B}^{(2)}$ and 
$F^{(1)} = 2 dA^{(1)}$ are the expressions for the curvatures
in the absence of the NS/NS 2--form gauge field.

We should point out a few subtleties in obtaining the expressions
(\ref{RRgf}). Starting from the particle, a straightforward
 application of the duality rules (\ref{duality2}) leads one to the
3-brane. Note that all $p$--brane solutions with $0\le p \le 2$
are electrically charged. In going from the 2-brane to the 3-brane
we dualize over the $x^3$ direction and apply the first rule of 
(\ref{duality2}) to obtain the
expression for $D_{0123}$. The expression for $F(D)_{ijklm}$ is obtained 
by using the selfduality relation (\ref{selfduality}).
Note that the 3-brane solution is dyonic: it carries both 
electric and magnetic charge. Next, in going from the 3--brane
to the 4--brane solution we cannot apply the first rule of
(\ref{duality2}) the reason being that the duality is now performed
over $x^4$. Instead, we first write $C_{ijk} = 8/3D_{ijk4}$ and take the
curl with respect to $x^l$ which gives $G_{ijkl} = 10/3 F(D)_{ijkl4}$.
We next substitute for $F(D)_{ijkl4}$ the expression corresponding
 to the 3--brane solution.
We thus obtain the expression for $G_{ijkl}$ given in (\ref{RRgf}).
Note that this 4-brane is magnetically charged. The expressions for
 the magnetically charged 5--brane and 6--brane solutions
are 
obtained by taking curls of the duality relations (\ref{duality2}).
The same applies to the transition from the 6--brane to 7--brane
solution where we dualize over the $x^7$ direction,
but note that by taking the curl of $\ell$ the $m$--dependent term
in the duality rule drops out since we take the derivative
with respect to $x^i\ (i=8,9)$ of the equation $\ell = A_7^{(1)} + 
m x^7$. Having arrived at the 7--brane one cannot repeat the 
process again. The reason for this is that the
independence of the harmonic function of the 7--brane solution of one of
the two transverse directions, say $x^8$, does {\it not}
imply that the RR scalar $\ell$ is independent of $x^8$.
In fact, in this case the expression for $\ell$ is given by
\cite{Be1}

\begin{equation}
\ell = \bigl (\partial_9 H \bigr )x^8\, ,
\end{equation}
where $x^9$ refers to the second transverse direction.
To go to the 8--brane we now perform duality over the $x^8$ direction 
and use the fact that in the massive $T$--duality rules given in 
(\ref{duality2}) the RR scalar $\ell$ is allowed to depend linearly on the
duality direction $x^8$. This leads to the 8-brane solution
 of \cite{Wi2}.
Finally, in the last step, we obtain 10--dimensional
Minkowski spacetime as the 9--brane solution. 

\vspace{.5cm}

\noindent{\bf 3. $D$--Brane Actions}

\vspace{.5cm}

Having established the $T$--duality between the Dirichlet $p$--brane solutions
 of IIA/IIB supergravity, 
 we next turn our attention to the 
 corresponding $D$--brane actions. Since the $D$--brane
 actions should provide for the source terms of the Dirichlet $p$--brane
 solutions we must be able to establish a duality between the 
 $D$--brane actions as well. 

In the recent literature \cite{To2,Sc1,Al1,Ts1,Gr1} the structure
 of the (bosonic part of the) $D$--brane actions for $0\le p \le 3$
 has been established.
It turns out that the coupling to the NS/NS background fields 
(i.e. the metric, dilaton and 2--form gauge field) is described
by a kinetic term of the Born--Infeld (BI) type \cite{Le1,Fr1}. 
The action contains besides the usual embedding coordinates $X^\mu$, an
additional worldvolume gauge field $V_i\ (i=1,\cdots p+1)$. For $p=1$
and flat background, such an action was proposed some time ago
in an attempt to
 give a spacetime scale--invariant formulation of the superstring \cite{Be3}.
In such a formulation the (dimensionful) string coupling constant arises as an
integration constant in solving the equation of motion of the worldvolume
 gauge field.
The $D$--brane actions contain also a Wess-Zumino (WZ) term that
 describes the coupling of the $D$--brane to the RR gauge fields
as well as further couplings to the NS/NS 2--form gauge field.

We will show that the structure of both the BI kinetic term
and the WZ term is determined by: (1) gauge 
invariance\footnote{One can even show that the RR gauge fields cannot
be coupled in a gauge--invariant way to the $D$--brane without the
worldvolume gauge field. 
This provides another explanation for the presence of this
gauge field.}
and (2) $T$-duality. In fact, as we will see below,
these requirements lead to extra terms
in the type IIA $D$--brane action in case the IIA supergravity background
contains a nonvanishing target space cosmological constant.

Since gauge symmetries play an important role in the present
 discussion, we first summarize the transformation rules of the
different RR gauge fields and of the worldvolume gauge field.
The (target space)
transformations of the IIA gauge fields are given by

\begin{eqnarray}
\label{gIIA}
\delta A^{(1)} &=& d\Lambda^{(1)} - {m\over
 2}\eta^{(1)}\, ,\nonumber\\
\delta B^{(1)} &=& d\eta^{(1)}\, ,\\
\delta C &=& d\chi + 2B^{(1)} d\Lambda^{(1)} - m\eta^{(1)} B^{(1)}\, ,
\nonumber
\end{eqnarray}
while those of the IIB gauge fields have the form

\begin{eqnarray}
\label{gIIB}
\delta {\cal B}^{(i)} &=& d\Sigma^{(i)}\, ,\hskip 1truecm i=1,2\, ,\\
\delta D &=& d\rho - {3\over 4} d\Sigma^{(1)}{\cal B}^{(2)}
+ {3\over 4}d\Sigma^{(2)} {\cal B}^{(1)}\, .\nonumber
\end{eqnarray}
The (target space)
transformation of the worldvolume gauge field is given by

\begin{eqnarray}
\delta V &=& {1\over 2}\eta^{(1)}\, ,\hskip 1.5truecm ({\rm IIA},\ p\ {\rm
 even})\, ,\nonumber\\
\label{gWV}
\delta V &=& {1\over 2}\Sigma^{(1)}\, ,\hskip 1.5truecm ({\rm IIB},\ p\ {\rm
 odd})\, ,
\end{eqnarray}
where the first rule is an abbreviation for $\delta V_i = 1/2 \partial_i X^\mu
\eta_\mu^{(1)}$.

As mentioned above, the $D$-brane action consists of two terms,
a kinetic and a WZ term:

\begin{equation}
S^{(p)} = \int d^{p+1}\xi \biggl ( {\cal L}^{(p)}_{{\rm kin}} +
{\cal L}^{(p)}_{{\rm WZ}} \biggr)\, .
\end{equation}
For all $p\ (0\le p\le 9)$ the kinetic term is of the following
BI-type:

\begin{equation}
\label{Lkin}
{\cal L}^{(p)}_{{\rm kin}} = e^{-\sigma}
\sqrt { |{\rm det} \bigl (g_{ij} + {\cal F}_{ij} \bigr )|}\, ,
\end{equation}
where $g_{ij}$ is the embedding metric:

\begin{eqnarray}
\label{emb}
g_{ij} &=& \partial_i X^\mu \partial_j X^\nu g_{\mu\nu}\, ,
\hskip 1.5truecm ({\rm IIA},\ p\ {\rm even})\, ,\nonumber\\
g_{ij} &=& \partial_i X^\mu \partial_j X^\nu j_{\mu\nu}\, ,
\hskip 1.5truecm ({\rm IIB},\ p\ {\rm odd})\, ,
\end{eqnarray}
and where $\sigma$ for $p$ even/odd is defined in 
eqs.~(\ref{sigma}). The (gauge--invariant) curvature ${\cal F}$ 
of the worldvolume gauge field is given by

\begin{eqnarray}
\label{curvature}
{\cal F} &=& 2dV - B^{(1)}\, ,\hskip 1.5truecm ({\rm IIA},\ p\ {\rm
 even})\, ,
\nonumber\\
{\cal F} &=& 2dV -{\cal B}^{(1)}\, ,\hskip 1.5truecm ({\rm IIB},\ p\ 
{\rm odd})\, ,
\end{eqnarray}
where the first line is short--hand for ${\cal F}_{ij} = 
\partial_i V_j - \partial_j V_i - \partial_i X^\mu \partial_j X^\nu
B_{\mu\nu}^{(1)}$.
The explicit dilaton coupling in front
of the kinetic term (\ref{Lkin}) corresponds to the fact that the mass per
unit $p$--volume of the
$D$ -brane scales as \cite{Wi1}

\begin{equation}
{\cal M}_p \sim {1\over g}\, ,
\end{equation}
where $g\sim e^\sigma$ is the string coupling constant.

Before giving the WZ terms we first discuss
 the $T$--duality of the kinetic terms. We first show, as an
example, how the duality works between the 0--brane and 1--brane action 
and next prove the duality for general $p$\footnote{For flat backgrounds,
the $T$--duality between the kinetic terms has been considered in \cite{Ba1}.
We thank P.~Townsend for bringing this reference to our attention. For
curved backgrounds the discussion below overlaps with the one given 
in \cite{Alv1}, see Note Added.}.
The kinetic terms are, respectively, given by\footnote{
Since we will show the duality by reduction to nine dimensions, it
is necessary at this stage to distinguish between
nine-- and ten--dimensional objects. In the discussion below
we will indicate the ten--dimensional indices with a hat.}:

\begin{equation}
{\cal L}^{(0)}_{{\rm kin}} = e^{-\hat\phi}
\sqrt { \dot {X}^{\hat\mu} \dot {X}^{\hat\nu} \hat
 g_{\hat\mu\hat\nu} }
\, ,
\end{equation}

\begin{equation}
{\cal L}^{(1)}_{{\rm kin}} = e^{-\hat \varphi} \sqrt {
|{\rm det } \bigl ({\hat g}_{ij} + {\hat {\cal F}}_{ij}\bigr)| }\, ,
\end{equation}
where the embedding metric $g_{ij}$ is defined by the second
 equation of (\ref{emb}).

We first consider the reduction of the 0--brane action to nine
 dimensions. Like in the
 previous section, we dualize over the $x^{\hat 1}$--direction. Since,
from the $0$--brane point of view, this is a so--called direct
 dimensional reduction, we obtain after reduction a
$0$--brane action in nine dimensions where the $x^{\hat 1}$ coordinate 
has become an extra worldline scalar $S$, i.e.\footnote{ 
This is similar to the
reduction of the eleven--dimensional membrane to ten dimensions
\cite{To2}. A difference is that in the reduction of the
 eleven--dimensional
membrane the extra scalar becomes, after dualization on the
 3--dimensional worldvolume, a worldvolume gauge field.
In the particle case the scalar remains a scalar. The role of this
extra scalar in nine dimensions
is however similar to the role of the worldvolume vector
in ten dimensions. In the same way as the worldvolume vector
is used to give a scale--invariant description of the string
 where the string tension arises as an integration constant,
the extra scalar in nine dimensions can be used to give a scale--invariant
formulation of the massive particle where the mass arises as an
 integration constant.
The massive particle in nine dimensions can then be interpreted as
a massless particle in ten dimensions. For more details, see \cite{Be3}.}

\begin{equation}
X^{\hat 1} = S\, .
\end{equation}
Using the reduction formulae given in \cite{Be1,Be2} the kinetic term
for the nine--dimensional 0--brane action, coupled to the
background fields of $D=9$ supergravity
$\{ g_{\mu\nu}, C_{\mu\nu\rho}, B_{\mu\nu}^i, A_\mu^i, 
B_\mu, \phi, k, \ell\}\ 
(i=1,2)$\footnote{The $D=9$ NS/NS fields are $g, B^{(1)}, 
A^{(2)}, B, \phi$ and $k$.}, is given by

\begin{equation}
\label{pd=9}
{\cal L}_{{\rm kin}}^{(0)} = 
e^{-\phi}k^{-{1\over 2}}\sqrt {
\dot X^\mu \dot X^\nu \bigl (g_{\mu\nu}-k^2A_\mu^{(2)}A_\nu^{(2)}\bigr )
-k^2 {\dot S}^2 -2 k^2 \dot S {\dot X}^\mu A_\mu^{(2)} }\, .
\end{equation}

We next consider the reduction to nine dimensions of the 1-brane
action. Again we reduce over the $\hat 1$--direction but in this case
we call the
corresponding embedding coordinate of the 1--brane
$Y^{\hat 1}$ to distinguish it from the
0--brane coordinate $X^{\hat 1}$ used above.
Since in this case we are performing a double dimensional
reduction we set

\begin{equation}
Y^{\hat 1} = \sigma\, ,
\end{equation}
where $\sigma$ is the spacelike direction of the 1--brane worldvolume.
It turns out that, in order to obtain {\it the same} nine-dimensional
0--brane action (\ref{pd=9}) we must compactify the worldvolume
gauge field as follows:

\begin{equation}
{\hat V}_{\sigma} = S\, .
\end{equation}
The time component of $V$ drops out of the action
after double dimensional reduction.

To work out the dimensional reduction of the 1--brane action 
it is useful to first use the fact that for a $2\times 2$ matrix 
we have the identity:

\begin{equation}
{\rm det} \bigl (g_{ij} + {\cal F}_{ij}\bigr ) = 
{\rm det} g_{ij} - {\rm det} {\cal F}_{ij}\, .
\end{equation}
We thus obtain that ${\cal L}_{{\rm kin}}^{(1)}$ reduces to
the following expression after dimensional reduction:

\begin{equation}
e^{-\phi}k^{{1\over 2}}\left [
{\rm det } \pmatrix{
{\dot X}^\mu{\dot X}^\nu ( g_{\mu\nu} - k^{-2}B_\mu B_\nu )&&
-k^{-2}{\dot X}^\mu B_\mu\cr
-k^{-2}B_\mu && - k^{-2}\cr
}
+ {\hat {\cal F}}_{01}^2 \right ]^{{1\over 2}}\, .
\end{equation}
All $B$--terms in the determinant cancel.
Using the fact that

\begin{equation}
{\hat {\cal F}}_{01} = {\dot S} + {\dot X}^\mu A_\mu^{(2)}\, ,
\end{equation}
it is then easy to show that indeed the 1-brane action reduces to
the same nine--dimensional 0--brane action (\ref{pd=9}), thereby
establishing the $T$--duality between the 0--brane and 1--brane action.

It is interesting to see that in order to establish $T$--duality
the worldvolume
gauge field component ${\hat V}_\sigma$ of the 1--brane
gets identified with the 0--brane embedding coordinate $X^{\hat 1}$
leading to the $T$-duality relation

\begin{equation}
X^{\hat 1} = {\hat V}_\sigma\, .
\end{equation}
Another noteworhy feature is that the $T$--duality only works
in the presence of the dilaton coupling $e^{-\sigma}$ in front of the
kinetic term. Thus, $T$--duality requires that the mass per unit
$p$ volume of the $D$--brane is proportional to the inverse
string coupling constant.

Let us now prove $T$--duality between the kinetic terms for general $p$. 
 We first consider $p$ even, 
 so that we establish the duality between a IIA $p$-brane and a 
 IIB $(p+1)$-brane. We establish duality by reducing both kinetic terms
 to $d=9$, by direct and double dimensional reduction respectively,
 and by then showing that the resulting kinetic terms are equal.
 
As for the 0--brane, the transverse coordinate $X^{{\hat {(p+1)}}}$
 of the IIA $p$--brane becomes a woldvolume scalar: $X^{{\hat {(p+1)}}} = S$,
 while the worldvolume vector reduces as ${\hat V}_i = V_i$.
 The kinetic term for the $p$--brane is then determined by the
 following $(p+1) \times (p+1)$--matrix ($i,j = 0,1,\ldots p$)
 in nine dimensions:
\begin{equation}
 {\cal A}_{ij} = g^{A}_{ij} +{\cal F}^{A}_{ij}
\end{equation}
with
\begin{eqnarray}
 g^{A}_{ij} &=& G_{ij} -k^2 \bigl( A^{(2)}_i + S_i\bigr)
 \bigl( A^{(2)}_j + S_j \bigr)
 \, ,\\
 {\cal F}^{A}_{ij} &=& F(V)_{ij} - B_{ij} + B_i\bigl(S_j +A^{(2)}_j\bigr) 
 - B_j\bigl(S_i +A^{(2)}_i\bigr) \, ,
\end{eqnarray}
where we have used the notation:
\begin{eqnarray}
 G_{ij} &=& \partial_i X^\mu \partial_j X^\nu g_{\mu\nu}\, ,\nonumber\\
 B_{ij} &=& \partial_i X^\mu \partial_j X^\nu\bigl(
 B_{\mu\nu}^{(1)} - A^{(2)}_{[\mu}B_{\nu]}\bigr)\, \nonumber\\
A^{(2)}_i &=& {\partial_i X}^\mu A_\mu^{(2)}\, ,\nonumber\\
 B_i &=& {\partial_i X}^\mu B_\mu\, ,\nonumber\\
 S_i &=& \partial_i S\, .
\end{eqnarray}

For the IIB $(p+1)$--brane we perform a double dimensional reduction.
 One of the worldvolume scalars is identified with a spacelike
 worldvolume coordinate: $X^{\hat{ (p+1)}} =\rho$, while the
 $(p+1)$-component of $V$ becomes a worldvolume scalar:
 ${\hat V}_{p+1} = S,\ {\hat V}_i = V_i$.
 The IIB $p+1$--brane then gives in nine dimensions the following
 $(p+2)\times (p+2)$--matrix:
\begin{equation}
 {\cal B} = g^B + {\cal F}^B\, ,
\end{equation}
with
\begin{eqnarray}
 g^{B} &=& \pmatrix{ G_{ij} - k^{-2}B_iB_j
 && -k^{-2}B_i\cr
 -k^2 B_j&& - k^{-2}\cr } \, ,\nonumber\\
 {\cal F}^B &=& \pmatrix{ F(V)_{ij}-B_{ij}
 && S_i+B_i\cr
 -(S_j+B_j)&&0 \cr} \, .
\end{eqnarray}
The definitions
\begin{eqnarray}
 U^{\pm}_i&\equiv& S_i + A^{(2)}_i \pm k^{-2}B_i\,,\nonumber\\
 H_{ij} &\equiv& G_{ij} +F(V)_{ij} -B_{ij} -k^{-2}B_iB_j\,,
\end{eqnarray}
lead to the following form of ${\cal A}$ and ${\cal B}$:
\begin{eqnarray}
 {\cal A}_{ij} &=& H_{ij} - k^2 U_i^-U_j^+\,,\nonumber\\
 {\cal B} &=&
 \pmatrix{ H_{ij} && U_i^-\cr -U^+_j && -k^{-2} \cr}\, .
\end{eqnarray}
The determinant of ${\cal B}$ can now be easily obtained.
We rewrite
\begin{equation}
 {\cal B} =
 \pmatrix{ H_{ij} && 0 \cr 0&&-k^{-2}}
 \pmatrix{ \delta_{kj} && H^{-1}_{kl}U_l^-\cr
 k^2U^+_j && 1 \cr} \,,
\end{equation}
and show by induction that
\begin{equation}
 \det {\cal B} = -k^{-2} \big(1 - k^2U^+_kH^{-1}_{kl}U^-_l)\det H\, .
\end{equation}
The matrix ${\cal A}$ we rewrite as
\begin{equation}
\label{AUU}
 {\cal A}_{ij} = H_{ik}\bigl( \delta_{kj} - k^2H^{-1}_{kl}U^-_lU^+_j
 \bigr)\, .
\end{equation}
The second factor in (\ref{AUU}) has $p$ eigenvalues equal to one (for the
 $p$ vectors orthogonal to $U^+$), and one eigenvalue equal to
 $1 - k^2H^{-1}_{kl}U^-_lU^+_k$, for the eigenvector
 $H^{-1}_{ij}U^-_j$. The calculation of the IIA,B determinants
 assumes that the inverse of $H$ exists, and that $H^{-1}_{ij}U^-_j$
 is not orthognal to $U^+_j$. This is clearly correct for generic
 field configurations.

The calculation of the determinants establishes the equality
\begin{equation}
 \det{\cal B} = -k^{-2}\det {\cal A}\,.
\end{equation}
The factor $k^{-2}$ is cancelled by the dilaton prefactor in the 
 D--brane action. So indeed we obtain by reduction to $d=9$ the same 
 kinetic terms from the IIA $p$--brane and the IIB $(p+1)$--brane.

In case $p$ is odd, the roles of the IIA and IIB fields are interchanged.
 The same argument as above, with the interchange of the $d=9$ fields
 $A^{(2)} \leftrightarrow B$ and $k \leftrightarrow k^{-1}$
 leads to the proof of duality for $p$ odd.

Having established the duality for the kinetic terms, we now turn our 
 attention to the Wess-Zumino terms. These
 terms have been studied in \cite{To2,Sc1,Al1,Ts1,Gr1}.
Most of the contributions to the Wess--Zumino term are fixed by the
gauge symmetries given in (\ref{gIIA}), (\ref{gIIB}) and (\ref{gWV}).
We obtain the following result for $0\le p\le 3$ (omitting hats and
using form notation):

\begin{eqnarray}
{\cal L}_{{\rm WZ}}^{(0)} &=& A^{(1)} +
 mV_t\, ,\nonumber\\
{\cal L}_{{\rm WZ}}^{(1)} &=& {1\over 2} {\cal B}^{(2)} + {1\over 2}
\ell {\cal F}\, ,\nonumber\\
{\cal L}_{{\rm WZ}}^{(2)} &=& {1\over 4} C + {1\over 2} A^{(1)}{\cal F}
+ {m\over 2} VdV\, ,\\
{\cal L}_{{\rm WZ}}^{(3)} &=& {1\over 6} D +{1\over 4}{\cal
 B}^{(2)}{\cal F} +{1\over 8} {\cal B}^{(1)}{\cal B}^{(2)}
+ {1\over 16}\ell {\cal F}^2\, .\nonumber
\end{eqnarray}
Note that since the target spacetime gauge transformations contain 
$m$--de\-pen\-dent
terms, gauge invariance requires, for even $p= 0,2$, 
the presence of 
$m$--dependent terms in the Wess-Zumino term.
For $p=0$ the equation of motion of the worldline gauge field $V_t$
(which is absent in the $p=0$ kinetic term)
leads to $m=0$. Finally, we note that the 3--brane action
contains an explicit ${\cal B}^{(1)}{\cal B}^{(2)}$ term
which arises as a higher order term in string theory
(see footnote 10 of \cite{Ts1}).

We have explicitly verified that the above WZ terms are
 not only gauge--invariant but also $T$--dual to each other. 
 The requirement of duality fixes the coefficients of
 the $\ell {\cal F}^{(p+1)/2}$ terms in the 1--brane and 3--brane action
 which are gauge--invariant by themselves. As an illustration we
 show how the duality between the $p=0$ and $p=1$ WZ
 terms is established. Introducing the hat notation and taking 

\begin{equation}
X^{\hat 1} = S\, \hskip 1.5truecm {\hat V}_t = V_t\, ,
\end{equation}
the $p=0$ WZ term reduces to

\begin{equation}
\label{WZd=9}
{\dot X}^\mu\bigl (
A_\mu^{(1)} + \ell A_\mu^{(2)}\bigr ) + \ell \dot S + mV_t\, .
\end{equation}
Similarly, taking 

\begin{equation}
Y^{\hat 1} = \sigma\, ,\hskip 1truecm {\hat V}_\sigma = S\, ,
\hskip 1truecm {\hat V}_t = V_t\, ,
\end{equation}
the $p=1$ WZ term first reduces to\footnote{
At first sight the term $\partial_\sigma V_t$ might seem odd
since $V_t$ is independent of $\sigma$ and hence this term would
seem to vanish. However, our rule of dimensional reduction is that
one should first, by partial differentiation, eliminate
 the $\sigma$--dependence from the action. This same ambiguity arises
in the dimensional reduction of IIB supergravity discussed in
\cite{Be1}.}

\begin{equation}
{\dot X}^\mu\bigl (A_\mu^{(1)} - m\sigma A_\mu^{(2)}\bigr )
+ \bigl (\ell + m\sigma\bigr )\bigl ({\dot S} -\partial_\sigma V_t
+ {\dot X}^\mu A_\mu^{(2)}\bigr )\, .
\end{equation}
After partial differentiation the $\sigma$--dependent terms cancel
and one is left with the same expression (\ref{WZd=9}). 
This establishes the $T$--duality. The
 cancellation of the $\sigma$--dependent terms is related to
the fact that we are performing a Scherk--Schwarz dimensional
reduction \cite{Scherk1} that makes use of a global $U(1)$ subgroup
of $SL(2,R)_{{\rm IIB}}$, as is explained in \cite{Be1}. Note that
 the duality establishes a relation between the gauge--invariant
$\ell{\cal F}$ term in the $p=1$ WZ term with the $mV_t$
topological term in the $p=0$ WZ term whose coefficient
is determined by gauge invariance. It is in this way that we
have used $T$--duality to determine the coefficients of the
gauge--invariant $\ell {\cal F}^p$ terms in the $p=1$ and $p=3$
WZ terms.

Finally, we note that the same $p=1$ WZ term also plays a
 role in establishing a duality between the $p=1$ and $p=2$
 WZ terms. However, in this case one should reduce the
 $p=1$ WZ term via a direct as opposed to a
 double dimensional reduction. 
 Assuming that we reduce over the 
 $\hat 2$--direction, the $p=1$ WZ term is reduced according to

\begin{equation}
Y^{\hat 2} = S\, ,\hskip 1truecm {\hat V}_{i} = V_i\, ,
\hskip 1truecm i=0,1\, .
\end{equation}
On the other hand, the $p=2$ WZ term is reduced by double dimensional
 reduction as follows:

\begin{equation}
X^{\hat 2} = \rho\, ,\hskip 1truecm {\hat V}_{\rho} = S\, ,
\hskip 1truecm {\hat V}_i = V_i\, ,\hskip 1truecm i=0,1\, ,
\end{equation}
 where $\rho$ is a spacelike direction on the $p=2$ worldvolume.
 The same identifications were used in establishing the duality
 between the kinetic terms.

\vspace{.5cm}

\noindent{\bf 4. Comments}

\vspace{.5cm}

In this letter we have shown how 
 $T$--duality is realized on (i) the Dirichlet $p$--brane
 solutions of IIA/IIB supergravity for $0\le p\le 9$, (ii) the kinetic 
 terms of the $D$--brane
 actions for $0\le p\le 9$ and (iii) the WZ terms of the $D$--brane 
 actions for $0\le p\le 3$. To establish $T$--duality 
 between the WZ terms for $3\le p
 \le 9$ we first need to know their form. Note that for these
 cases there is no leading order RR gauge field in the formulation
 of IIA/IIB supergravity we are using here
 that naturally couples to the
 $D$--brane. Such higher--order RR gauge fields can be obtained 
 from the existing ones by dualization but sofar this seems to
 lead to rather complicated expressions. 
 
We find that the $p=0,2$ WZ terms contain $m$--dependent topological
terms for the worldvolume gauge field. For $p=0$ the 
field equation of the worldvolume gauge field leads to $m=0$. Note
that only for $m=0$ the 0--branes can be interpreted as KK states
of $D=11$ supergravity \cite{To1}. Concerning the $p=2$ WZ term
we note that only for $m=0$ 
the 2--brane action can be obtained by direct
dimensional reduction of the eleven-dimensional supermembrane \cite{To2}.
This 11--dimensional interpretation is not possible
 for $m\ne 0$ since in that
 case the worldvolume gauge field $V$ cannot be dualized into a
 scalar due to the topological mass term.
This is related to the fact that massive
IIA supergravity has no eleven--dimensional interpretation, at least
not that we know of. 
 
Finally, our hope is that the results of the present work will be useful to
establish a kappa--symmetric extension of the bosonic $D$--brane actions. 
Such a kappa--symmetric extension has been constructed
for the 0-brane and for the 2--brane with $m=0$ \cite{To2}. The 
$T$--duality relations given here should be helpful in 
constructing
kappa--symmetric extensions of the other $D$--brane actions. A
particularly interesting case is the 9--brane action which in the 
``physical gauge'' becomes equal to a supersymmetric BI action. 
We hope to report 
on progress in this direction in a future publication. 

\vspace{.5cm}

\noindent NOTE ADDED:\ \ Upon completing this work we received a preprint
\cite{Alv1} which also discusses the $T$--duality between $D$--brane actions.

\vspace{.5cm}

\noindent {\bf Acknowledgements}
\vspace{.5truecm}

We thank G.~Papadopoulos, T.~Ort\'\i n and P.~Townsend for useful discussions.
The work of E.B.~has been made possible by a fellowship of the Royal 
Netherlands Academy of Arts and Sciences (KNAW).


\vspace{.5truecm}


\begin{thebibliography}{99}
\bibitem{Gi1} For a review, see A.~Giveon, M.~Porrati and E.~Rabinovici, 
Phys.~Rep.~{\bf 244} (1994) 77.
\bibitem{Pol1} J.~Polchinski, Phys.~Rev.~Lett.~{\bf 75} (1995) 184.
\bibitem{Pol2} J.~Polchinski, S.~Chaudhuri and C.V.~Johnson,
{\tt hep-th/9602052}.
\bibitem{Be1} E.~Bergshoeff, M.~de Roo, M.B.~Green, G.~Papadopoulos
and P.K.~Townsend, {\it Duality of Type II 7--branes and 8--branes},
{\tt hep-th/9601150}.
\bibitem{Be2} E.~Bergshoeff, C.M.~Hull and T.~Ort\'\i n, 
Nucl.~Phys.~{\bf B451} (1995) 547.
\bibitem{Da1} J.~Dai, R.G.~Leigh and J.~Polchinski, Mod.~Phys.~Lett.~A, Vol.~4,
No.~{\bf 21} (1989) 2073.
\bibitem{Di1} M.~Dine, P.~Huet and N.~Seiberg, 
Nucl.~Phys.~{\bf B322} (1989) 301.
\bibitem{To1} P.K. Townsend, Phys.~Lett.~{\bf 350B} (1995) 184.
\bibitem{Ro1} L.J.~Romans, Phys.~Lett.~{\bf 169B} (1986) 374.
\bibitem{Wi2}E.~Witten and J.~Polchinski, Nucl.~Phys.~{\bf B460} (1996) 525.
\bibitem{To2} P.K.~Townsend, {\it $D$--branes from $M$--branes},
{\tt hep-th/9512062}.
\bibitem{Sc1} C.~Schmidhuber, {\it $D$--brane actions}, 
{\tt hep-th/9601003}.
\bibitem{Al1} S.P.~de Alwis and K.~Sato, {\it $D$--strings
 and $F$--strings from string loops}, {\tt hep-th/9601167}.
\bibitem{Ts1} A.A.~Tseytlin, {\it Self--duality of Born--Infeld action
and Dirichlet 3--brane of type $IIB$ superstring theory},
 {\tt hep-th/9602064}.
\bibitem{Gr1} M.B.~Green and M.~Gutperle, {\it Comments on
 three-branes}, {\tt hep-th/9602077}.
\bibitem{Le1} R.~Leigh, Mod.~Phys.~Lett.~{\bf A4} (1989) 2767.
\bibitem{Fr1} E.S.~Fradkin and A.A.~Tseytlin, 
Phys.~Lett.~{\bf B163} (1985) 123.
\bibitem{Be3} E.~Bergshoeff, L.A.J.~London and P.K.~Townsend, 
Class.~Quantum Gravity {\bf 9} (1992) 2545.
\bibitem{Wi1} E.~Witten, Nucl.~Phys.~{\bf B443} (1995) 85.
\bibitem{Ba1} C.~Bachas, {\it $D$--brane dynamics}, {\tt hep-th/9511043}.
\bibitem{Scherk1} J.~Scherk and J.H.~Schwarz, Phys.~Lett.~{\bf 82B} (1979)
60.
\bibitem{Alv1} E.~Alvarez, J.L.F.~Barb\'on and J.~Borlaf, {\it
$T$--duality for open strings}, {\tt hep-th/9603089}.
\end{thebibliography}
\end{document}